\documentclass{optica-article}

\journal{opticajournal} 

\articletype{Research Article}

\usepackage{lineno}
\usepackage{siunitx}

\begin{document}

\title{Fiber-based electro-optic dual-comb light source for fast linear and nonlinear spectroscopy}


\author{Debanuj Chatterjee\authormark{1*}$^\dagger$, Fabio Motetta\authormark{1,2}$^\dagger$, Simon Boivinet\authormark{1}, Herv\'e Rigneault\authormark{2} and Arnaud Mussot\authormark{1,3*}}

\address{\authormark{1}University of Lille, CNRS, UMR 8523 PhLAM (Physique des Lasers Atomes et Molecules), Lille, France}
\address{\authormark{2}Aix Marseille University, CNRS, Centrale Med, Institut Fresnel, Marseille, France}
\address{\authormark{3}Institut Universitaire de France (IUF), Paris, France.}
 \address{$^\dagger$ These authors contributed equally}
\email{\authormark{*}debanuj.chatterjee@univ-lille.fr}
\email{\authormark{*}arnaud.mussot@univ-lille.fr}


\begin{abstract} 
Dual-comb spectroscopy\,(DCS) enables {\color{black}rapid,} broadband and high-resolution optical measurements by mapping optical spectra into the radio frequency\,(RF) domain. However, conventional DCS systems are fundamentally constrained by a tradeoff between optical bandwidth and interferogram acquisition speed, limiting their overall performance. Here, we demonstrate an all-fiber, polarization-maintaining\,(PM) frequency-agile electro-optic modulation\,(EOM) dual-comb source that simultaneously achieves a broad optical bandwidth of 10 THz and a high interferogram acquisition speed of up to 2.5 MHz. The high acquisition rate is enabled through an in-phase/quadrature\,(IQ) modulator-based architecture to shift the carrier frequency of one of the combs. We illustrate the performance of the source through proof-of-concept linear spectroscopy and nonlinear dual-comb coherent anti-Stokes Raman scattering\,(CARS) spectroscopy measurements. The combination of large spectral coverage, high refresh rate and an all-PM fiber configuration makes this dual-comb platform attractive for applications such as rapid molecular spectroscopy and high-speed nonlinear spectroscopic imaging.
\end{abstract}

\section{Introduction}\label{secIntro}

Dual-comb spectroscopy\,(DCS) is an advanced optical sensing technique that enables rapid, high-resolution, and broadband spectral measurements without moving mechanical components\,\cite{coddington_dual-comb_2016,ideguchi2017dual}. It employs two optical frequency combs with slightly different repetition rates, $f_1$ and $f_2=f_1-\Delta f$ (where $\Delta f\ll f_1$ is the repetition rate difference between the combs), generating a radio frequency\,(RF) interferogram signal with a $\Delta f$ repetition rate, when the combs beat on a photodetector\,(PD). The down-converted interferogram signal maps optical frequencies into the RF domain, offering fast acquisition speeds and high sensitivity, making it valuable for fast molecular spectroscopy\,\cite{long2014multiheterodyne,ycas2018high,yan2017mid}, real-time atmospheric monitoring\,\cite{rieker2014frequency,zhu2015mid,coburn2018regional}, precise distance metrology\,\cite{nurnberg2021dual, wright2021two, chang2024dispersive}, to name a few. 

In a DCS system, an important parameter is the magnification factor $m=\frac{f_1}{\Delta f}$, which also characterizes the ratio between the spectral width of the optical combs to the detected down-converted RF comb. {\color{black}Practically, to avoid spectral aliasing}, $m$ must be large enough to compress the optical spectrum with a bandwidth $\Delta \nu_{opt}$, into a RF spectrum with bandwidth $\Delta \nu_{RF}=\frac{\Delta \nu_{opt}}{m}$ that fits within the frequency range between 0 and $\frac{f_1}{2}$\,\cite{coddington_dual-comb_2016}. This inherently creates a tradeoff between the spectral bandwidth $\Delta\nu_{opt}$ and interferogram acquisition speed $\Delta f$\,\cite{carlson2020broadband, voumard20221}. {\color{black}However, DCS systems benefit from both, a large spectral bandwidth (for detecting multiple spectral features over a large bandwidth simultaneously), and a large interferogram acquisition rate (enabling faster spectroscopy).}
Although the tradeoff between $\Delta\nu_{opt}$ and $\Delta f$ could be by-passed to some degree using time-programming of the pulse trains\,\cite{schliesser2005frequency,giorgetta2024broadband,caldwell2022time,chatterjee2025sensitivity,kameyama2020dual}, or spectral filtering of the combs\,\cite{nishiyama2017sensitivity,hoghooghi202111}, or using delay lines\,\cite{yu2021phase}, such techniques involve complicated implementation and/or induce extra constraints such as moving mechanical parts\,\cite{schliesser2005frequency, giorgetta2024broadband, caldwell2022time, kameyama2020dual} and higher optical losses\,\cite{nishiyama2017sensitivity,hoghooghi_broadband_2022,yu2021phase}. Thus here we limit our discussion to traditional DCS systems without advanced temporal or spectral modification of the combs. 

{\color{black}In order to reward higher acquisition rates along with larger spectral coverage, we define a figure of merit\,(FOM$=\Delta\nu_{opt}\times\Delta f$) for alias-free DCS systems. Note that here we neglected the contribution of spectral resolution in the FOM definition considering the comb resolution given by $f_1$ is much smaller than the linewidth of the target spectral feature we want to interrogate.}
Previous DCS works such as Okubo \emph{et al.} have reported a large spectral coverage of 140 THz, whilst being limited by an acquisition rate of $\Delta f=7.6$ Hz (FOM$=0.01$ GHz$^2$)\,\cite{okubo2015ultra}. On the other extreme, Long \emph{et al.} recently demonstrated a high acquisition rate of $\Delta f=150$ MHz, but with a low spectral coverage of 38 GHz (FOM$=5.7$ GHz$^2$)\,\cite{long_nanosecond_2024}. For reference, a non-exhaustive list of works on electro-optic modulation\,(EOM) comb-based DCS without aliasing, having a large value of FOM is shown in Table\,\ref{tab}. To the best of our knowledge, a DCS system with a large optical bandwidth, as well as a high acquisition rate (FOM$>20$ GHz$^2$) is unreported in the literature. In this work, we bridge that gap by demonstrating an all-fiber frequency-agile DCS setup capable of spectroscopy over a large optical bandwidth of about 10 THz, and at a high interferogram acquisition rate of 2.5 MHz, leading to a high value of FOM$=25$ GHz$^2$. To achieve a wideband operation, we developed two copies (for the dual-comb) of an all-fiber, polarization-maintaining\,(PM) EOM-based frequency comb delivering 60 fs pulses at 10 GHz repetition rate, similar to that reported in \cite{boivinet2025polarization}. While the EOM-based operation gave access to repetition rate tunability and high frequency operation ($f_1\sim$10 GHz)\,\cite{parriaux2020electro}, in order to operate at high acquisition speeds (or large $\Delta f$) without aliasing, we used an in-phase/quadrature\,(IQ) modulator-based architecture. The IQ modulator\,(IQM) imparts a tunable constant frequency shift to the continuous wave\,(CW) light in one of the combs in the GHz range, such that the down-converted RF comb is centered near the optimal RF frequency, \emph{i.e.} $\frac{f_1}{4}$, considering the RF comb extends from 0 to $\frac{f_1}{2}$ frequency.
{\begin{table}[h!]
    \centering 
    \begin{tabular}{|m{18em}|m{5.1em}|m{4.1em}|m{5.9em}|}
    \hline\centering
        \textbf{Paper} & $\mathbf{\Delta\nu_{opt}}$ \textbf{(GHz)}&$\mathbf{\Delta f}$ \textbf{(GHz)} &$\mathbf{FOM}$ \textbf{(GHz$^2$)} \\ \hline \hline
        Carlson \emph{et al.}, \emph{Optics Express} (2020)\,\cite{carlson2020broadband}&120$\times10^3$&10$\times10^{-6}$&1.2\\ \hline
        Long \emph{et al.}, \emph{Nature Photonics} (2024)\,\cite{long_nanosecond_2024}&38&150$\times10^{-3}$&5.7\\ \hline
        Chatterjee \emph{et al.}, \emph{IOP J. Phys. B} (2025)\,\cite{chatterjee2025real}&300&10$\times10^{-3}$&3\\ \hline
         This work&10$\times10^3$&2.5$\times10^{-3}$&25\\ \hline
    \end{tabular}
    \caption{Table for comparison of figure of merit\,(FOM$=\Delta\nu_{opt}\Delta f$) in different works on EOM comb-based alias-free DCS system.}
    \label{tab}
\end{table}}

Apart from linear spectroscopy, another important application avenue of dual-comb light sources lies in nonlinear spectroscopy techniques, such as dual-comb coherent anti-Stokes Raman scattering\,(CARS)\,\cite{ideguchi2013coherent,mohler2017dual,coluccelli2017fiber,carlson2020broadband}. {\color{black}In that context, our dual comb light source offers several advantages. For example, compared to Ti:sapphire mode-locked laser\,(MLL) based dual comb CARS setups\,\cite{ideguchi2013coherent, mohler2017dual}, EOM architecture is simpler to operate and can offer a more compact design. Furthermore, {\color{black}owing to the high repetition rate of the combs,} our system can achieve a high interferogram acquisition speed, order of magnitude higher than the state-of-the-art dual comb CARS systems, without any time-programming of the pulses\,\cite{carlson2020broadband, coluccelli2017fiber}.} 

To demonstrate the efficacy of our light source, we retrieved the low-frequency ($<100$ cm$^{-1}$) Raman active modes of Bi$_{12}$GeO$_{20}$\,(BGO) utilizing the dual-comb CARS technique. Thanks to the Fourier transform limited 60 fs pulses, the combs lead to an impulsive stimulated Raman scattering\,(ISRS) in the sample\,\cite{merlin1997generating}, leading to the dual-comb detection of the Raman frequency shift, through the down-conversion of the combs generated from the anti-Stokes photons. Furthermore, the high interferogram acquisition rate of our system allowed for fast dual-comb CARS spectroscopy, implying potential applications in CARS-based label-free imaging systems\,\cite{liu2026physics}.

\section{Dual-Comb Source}\label{secDC}

{\color{black}In this work, we extend our previously demonstrated single-comb source\,\cite{boivinet2025polarization} to a dual-comb architecture by implementing a second, nearly identical comb arm with a slightly different repetition rate, and combining the two outputs using an all-PM fiber combiner.}

\subsection{Experimental Setup}
\label{sub_setup}
\begin{figure}[htb!]
\centering
\includegraphics[width=1\columnwidth]{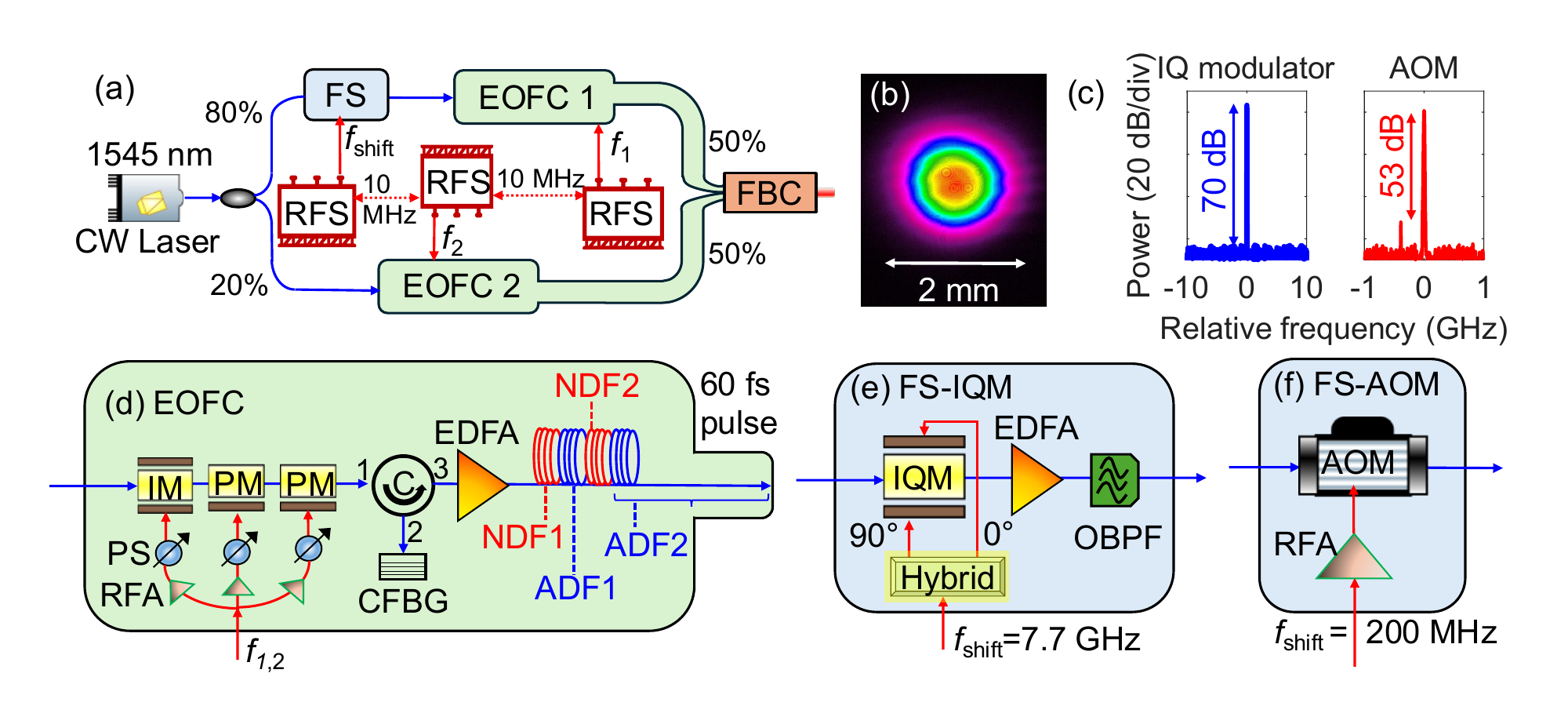}
\caption{(a) Experimental setup of dual-comb spectroscopy source. CW: Continuous Wave, FS: Frequency Shifter, RFS: Radio Frequency Synthesizer, EOFC: Electro-Optic Frequency Comb, FBC: Fiber Beam Combiner, $\Delta f$: repetition rate difference. {\color{black}EOFC is enumerated 1 and 2 corresponding to the two arms.} (b) False color plot of spatial beam profile recorded at a distance of 26 cm from the FBC output. (c) Extinction ratio comparison between acousto-optic modulator\,(AOM) and IQ modulator\,(IQM). (d) Electro-optic frequency comb scheme, IM: Intensity Modulator, PM: Phase Modulator, RFA: Radio Frequency Amplifier, PS: Phase Shifter, C: Circulator, CFBG: Chirped Fiber Bragg Grating, EDFA: Erbium-Doped Fiber Amplifier, NDF: Normal Dispersion Fiber, ADF: Anomalous dispersion fiber. {\color{black}NDF and ADF are enumerated 1 and 2 corresponding to the first and second segments.} (e) Frequency shifter implemented with IQM, Hybrid: 90° Hybrid coupler, OBPF: Optical Band-Pass Filter. (f) Frequency shifter implemented with AOM.}
\label{fig_setup}
\end{figure}

The experimental setup of the all-fiber, PM electro-optic dual-comb light source is schematically illustrated in Figs.\,\ref{fig_setup}\,(a,d,e,f). 
To generate the two combs, a shared low-linewidth ($\sim$100 Hz) CW laser from NKT Photonics at 1545 nm wavelength was used. The CW source was then split into two arms and each arm was sent through a cascade of electro-optic modulators (one intensity modulator and two phase modulators in series) driven at 10 GHz and 10 GHz-$\Delta f$, to generate two chirped electro-optic frequency combs in the two arms. $\Delta f$ was a tunable parameter with values ranging between 500 kHz to 2.5 MHz. The EOMs in the two arms were driven by two ultra-low noise radio frequency synthesizers\,(RFS) from Keysight (E8257D PSG). Note that the CW light in one of the arms was frequency shifted before the EO modulator stage, such that the down-converted RF interferogram is centered at that shifted frequency denoted as $f_{\text{shift}}$. Details on the frequency shifting will be discussed subsequently in Subsection\,\ref{subsec:AOMIQ}. 

\begin{figure}[htb!]
\centering
\includegraphics[width=1\columnwidth]{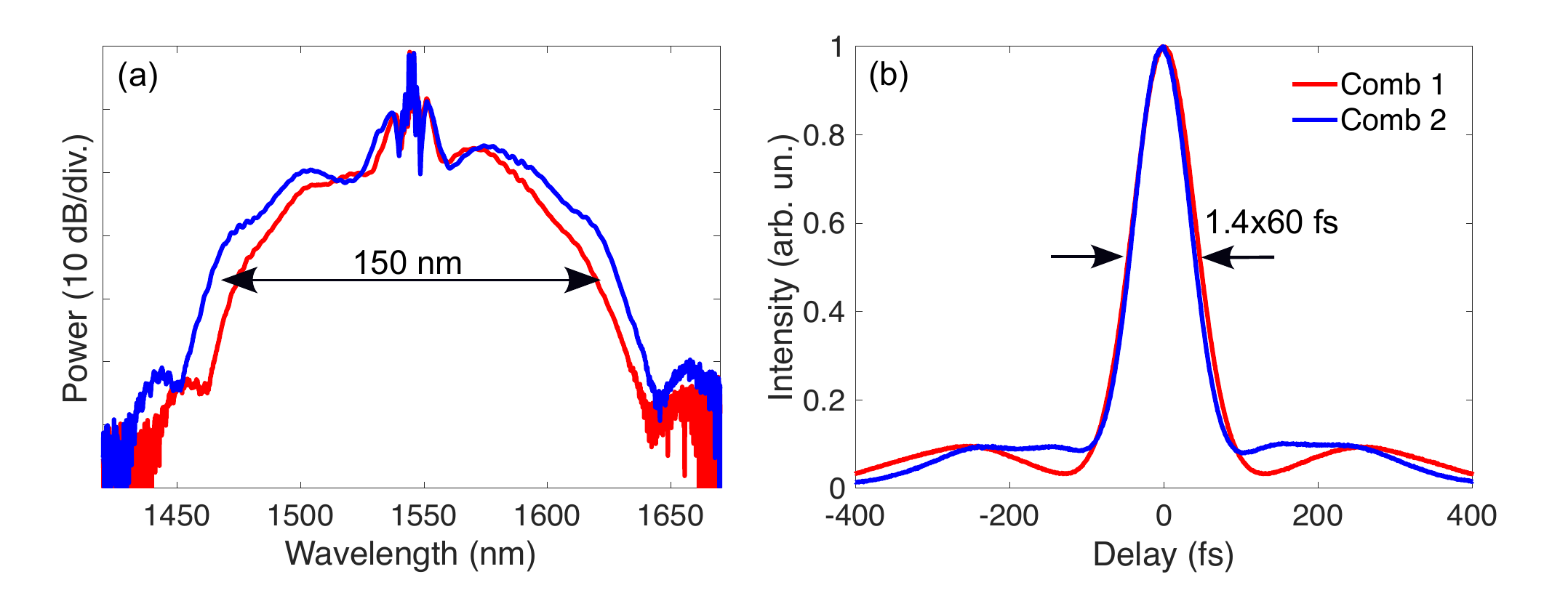}
\caption{(a) Optical spectra  and (b) autocorrelation traces for combs 1 (red) and 2 (blue), recorded with an optical spectrum analyzer and an auto-correlator respectively.}
\label{fig_spectra}
\end{figure}

The outputs of the EOM stages are launched into a chirped fiber Bragg grating (CFBG) to compensate for the linear frequency chirp induced by the phase modulators, resulting in nearly transform-limited 3 ps pulses. These pulses are then amplified via a high-power Erbium doped fiber amplifier\,(EDFA) (average output power \SI{\sim7}{\watt}) and injected into a two-stage nonlinear pulse compressor composed of alternating segments of normal (Coractive SCF-UN-3/125-25-PM) and anomalous (Coractive PM1550) dispersion fibers with optimized lengths as described in \cite{boivinet2025polarization}. {\color{black}The final stage of each comb compressor is formed by the PM fiber pigtails of a high-power PM fiber beam combiner\,(FBC), which combines the two combs leading to a free space output with excellent spatial overlap of the two beams\,[see anomalous dispersion fiber 2\,(ADF2) in Fig.\,\ref{fig_setup}\,(d)]. The beams are collimated and exhibit a Gaussian-like transverse intensity profile as shown in Fig.\,\ref{fig_setup}\,(b). The lengths of these final fiber segments were carefully adjusted to optimize the pulse compression at the FBC output, yielding nearly transform-limited $\sim$60 fs pulses with an average power of $\sim$5 W and a 40 dB spectral bandwidth of $\sim$150 nm (see Fig.\,\ref{fig_spectra}).}

\begin{figure}[tb]
\centering
\includegraphics[width=0.9\columnwidth]{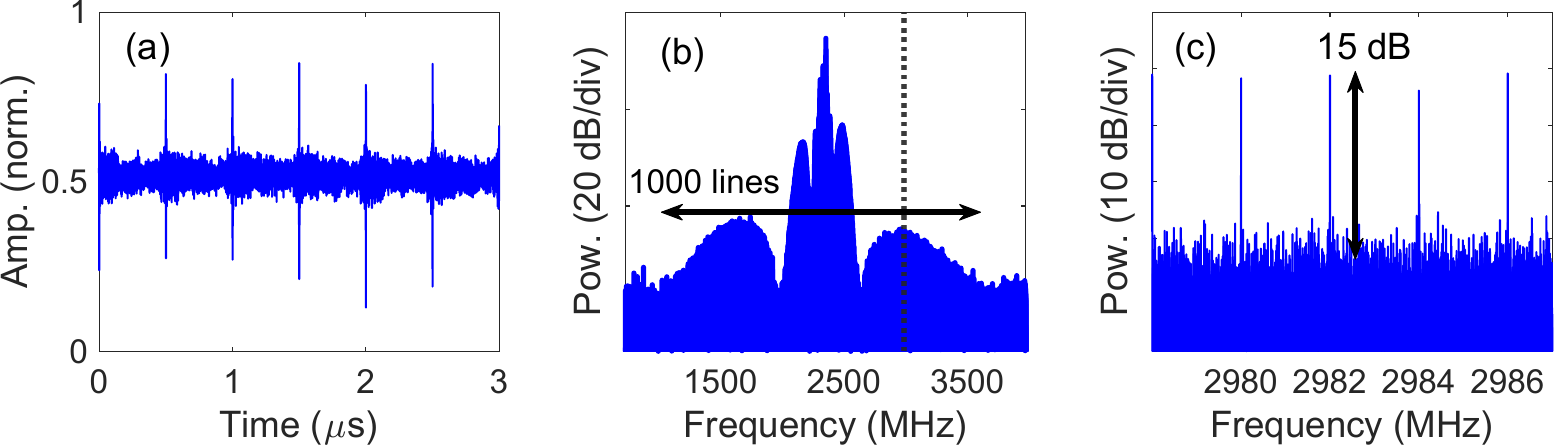}
\caption{Plot of (a) time-domain interferogram, (b) frequency-domain interferogram. (c) Zoom of (b) near 2982 MHz RF frequency [shown in dotted line in (b)].}
\label{fig_interferogram}
\end{figure}

The free-space dual-comb light after interacting with the sample (either through linear or nonlinear pathway) is launched onto a high-speed PD. The resulting RF signal is digitized in a 10 bit oscilloscope (Keysight Infiniium S-Series DSOS804A) to generate the final temporal interferogram. A sample interferogram recorded with a 5 GHz bandpass PD from Thorlabs (DETCFC08) is shown in Fig.\,\ref{fig_interferogram}. Fig.\,\ref{fig_interferogram}\,(a) shows the time domain interferogram with a 0.5 $\mu$s time period. On the other hand, Fig.\,\ref{fig_interferogram}\,(b) shows the frequency domain interferogram showing about 1000 comb lines. A zoomed portion of Fig.\,\ref{fig_interferogram}\,(b) is shown in Fig.\,\ref{fig_interferogram}\,(c) where we can clearly see individual mode resolved comb lines with a 15 dB signal-to-noise ratio\,(SNR).

\subsection{Dual Comb Characteristics}\label{subsec:AOMIQ}
{\color{black}
{\color{black}In this subsection we study the different characteristics of our dual comb setup. Firstly, in order to avoid spectral aliasing while maximizing the repetition-rate difference $\Delta f$, and hence minimizing the interferogram acquisition time, the down-converted RF comb should ideally be centered at $\frac{f_1}{4}$, \emph{i.e.}, at the center of the first Nyquist window extending from 0 to $\frac{f_1}{2}$\,\cite{coddington2008coherent,coddington_dual-comb_2016}. 
In our experiment, to maximize the extinction ratio of the IQM-based frequency shifter followed by spectral filtering [see Fig. 1(e)], we set $f_{\text{shift}} =7.7$ GHz, leading to a sideband suppression exceeding 70 dB\,[see left panel in Fig.\,\ref{fig_setup}\,(c), measured with a Brillouin optical spectrum analyzer\,(BOSA)]. A value of $f_{\text{shift}} =7.7$ GHz places the first-order RF comb at 2.3 GHz, close to the optimal value $\frac{f_1}{4}=2.5$ GHz. This near-optimal positioning of the RF comb enabled a repetition-rate difference as large as $\Delta f=2.5$ MHz without aliasing, corresponding to a minimum single-interferogram acquisition time of only 0.4 $\mu$s.}
Figure\,\ref{fig_snrIQ}\,(a) shows logarithmic scaling of the SNR of a RF comb line (10$^\text{th}$ line from the center for instance) as a function of the interferogram recording time $\tau=\frac{N_{\text{ifg}}}{\Delta f}$, where $N_{\text{ifg}}$ is the number of interferogram pulses in the acquired signal\,\cite{newbury_sensitivity_2010,duran2015ultrafast}. {\color{black}The SNR was calculated as the ratio of the power of the selected RF comb line to the mean noise power in an adjacent signal-free frequency interval [see Fig.\,\ref{fig_snrIQ}\,(b)].}
\begin{figure}[htb!]
    \centering
    \includegraphics[width=1\columnwidth]{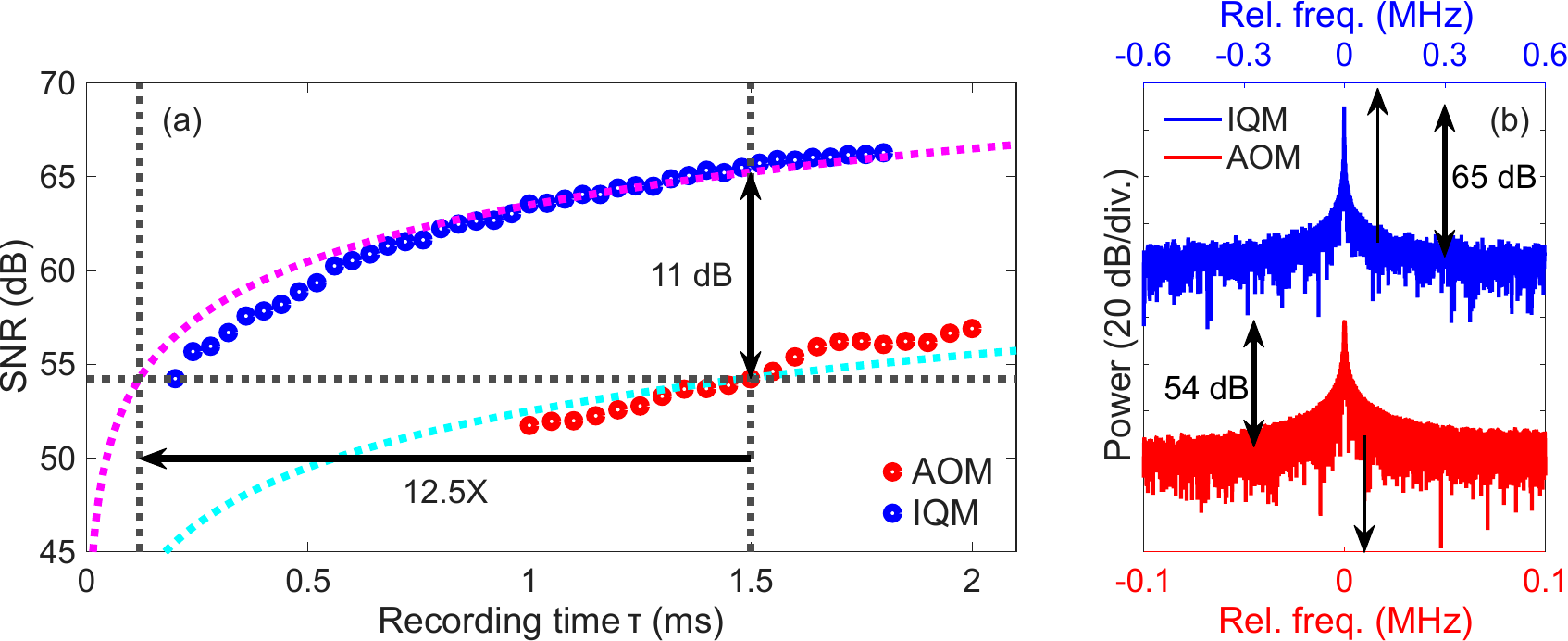}
    \caption{(a) Plot of SNR (of the 10$^{\text{th}}$ comb line) vs recording time $\tau$ for the IQM (blue) and acousto-optic modulator\,(AOM) based (red) configurations. The corresponding logarithmic fits [$\alpha+10\log_{10}(\tau)$] are shown in magenta and cyan dotted lines respectively. A time speedup of 12.5$\times$ for the IQM is shown for a given SNR of 54 dB. For a given recording time of 1.5 ms, the IQM attains 11 dB larger SNR shown in double-sided arrow. Black dotted lines are for visual aid. (b) Frequency domain interferogram showing the spectral region near the 10$^{\text{th}}$ comb line, for a 1.5 ms recording time, for IQM (blue) and AOM (red) cases. The curves are shifted vertically for visual clarity. The corresponding SNRs are shown with double sided arrows.}
    \label{fig_snrIQ}
\end{figure}

For comparison, we also implemented the frequency shift using an acousto-optic modulator\,(AOM), as conventionally employed in dual-comb spectroscopy systems\,[see Fig.\ref{fig_setup}\,(f)]\,\cite{millot2016frequency}. 
We used a 200 MHz AOM (Gooch \& Housego, Fibr-Q), corresponding to $f_{\mathrm{shift}}\ll\frac{f_1}{4}$.
The AOM provided approximately 53 dB optical extinction, compared with more than 70 dB obtained with the IQM configuration\,[see right panel in Fig.\,\ref{fig_setup}\,(c)]. {\color{black} In contrast to the IQM-based configuration, with the AOM-based configuration, avoiding aliasing required reducing $\Delta f$ to 200 kHz, corresponding to a minimum single-interferogram acquisition time of 5 $\mu$s, 12.5 times longer than with the IQM-based configuration. In other words, for a given recording time $\tau$, the IQM configuration can acquire a larger number of interferograms (or larger $N_{\text{ifg}}$) due to its larger $\Delta f$. This is also evident from Fig.\,\ref{fig_snrIQ}\,(a) showing a 11 dB higher SNR experimentally, in the IQM setup for a 1.5 ms recording time, close to the 10.9 dB expected gain from the larger number of interferograms acquired over the same time interval. Consequently, for a given SNR, the IQM case recording time is reduced by about 12.5 times compared to the AOM case.}
These measurements demonstrate that the frequency-agile IQM approach optimizes the placement of the down-converted RF comb within the available detection bandwidth, thereby enabling a larger attainable $\Delta f$. Rather than intrinsically increasing the SNR of the RF comb, this allows a larger number of pulses in the time domain interferogram to be acquired and coherently averaged within a fixed recording time, resulting in a substantial SNR and acquisition-speed advantage over AOM-based dual-comb systems.

}

\subsection{Phase Noise}
\begin{figure}[htb!]
\centering
\includegraphics[width=0.93\columnwidth]{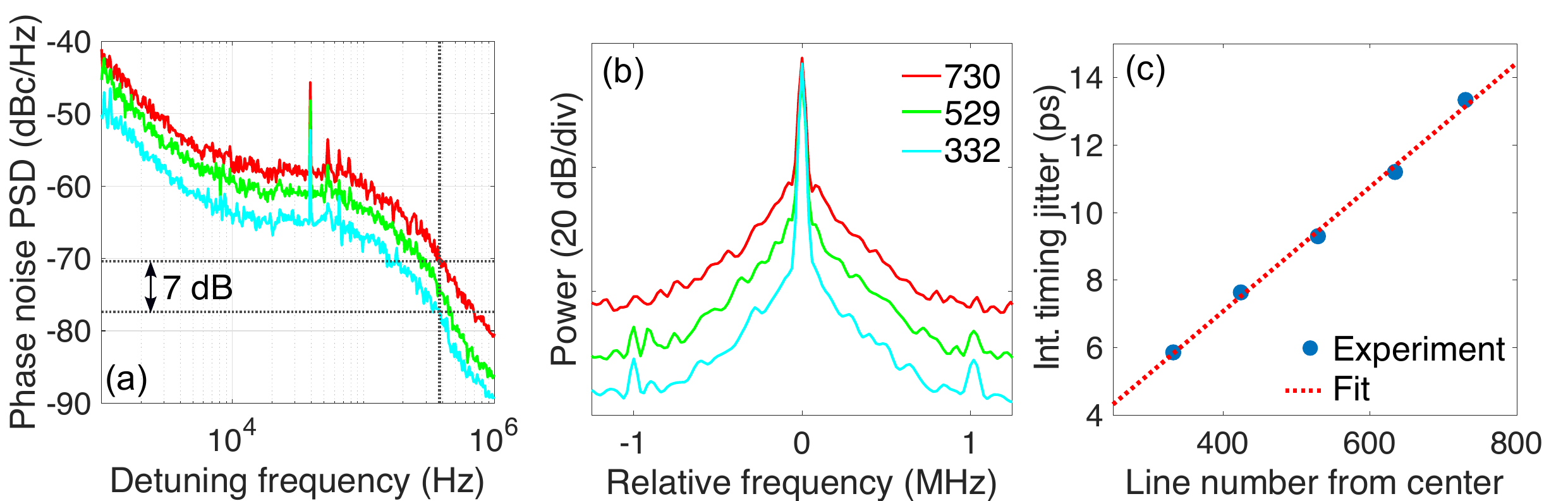}
\caption{Phase noise study. Plot of (a) single sideband phase noise power spectral density\,(PSD) vs detuning frequency and (b) electrical beatnote power spectrum of comb line numbers 332\,(cyan), 529\,(green) and 730\,(red) (from the comb center) obtained using an electrical spectrum analyzer\,(ESA). (c) Plot of integrated timing jitter (from 1 kHz to 1 MHz detuning) vs comb line number (blue filled circles). A linear fit is shown in red dotted line.}
\label{fig_phase_noise}
\end{figure}

The phase noise properties of the EOM combs, in the IQM-based configuration, were investigated by measuring the single sideband\,(SSB) phase noise power spectral density\,(PSD) of the beatnote between the $n^{\text{th}}$ comb line (from the comb center) of one comb and $(n+1)^{\text{th}}$ comb line of the other comb, with an electrical spectrum analyzer\,(ESA)\,\cite{chatterjee2025comparative}. It is well-known that the phase noise {\color{black}PSD} of the $n$-th comb line of a nonlinearly broadened EOM comb increases linearly with $n$\,\cite{ishizawa2013phase,chatterjee2025comparative,razumov2026impact,zeng2026integrated}. The phase noise PSD of the beatnote from two such combs scale as $n^2$\,\cite{chatterjee2025comparative}. We show in Fig.\,\ref{fig_phase_noise}\,(a) the SSB phase noise PSD between 1 kHz and 1 MHz detuning frequencies for comb line numbers 332\,(cyan), 529\,(green) and 730\,(red), corresponding to light around wavelengths 1519 nm, 1504 nm and 1489 nm respectively. The phase noise PSD for $n=730$ is about 7 dB higher than for $n=332$\,[see dotted line in Fig.\,\ref{fig_phase_noise}\,(a)], reinforcing the $n^2$ scaling that predicts $6.84$ dB\,\cite{chatterjee2025comparative}. {\color{black}The corresponding electrical spectra for the three lines $n=332, 529, 730$ are shown in Fig.\,\ref{fig_phase_noise}\,(b), with equalized peak powers. We clearly see that the beatnote pedestal power increases with increasing $n$.} Similarly, the integrated timing jitter (noise {\color{black}amplitude} integration between 1 kHz and 1 MHz detuning frequency) as a function of the comb line number $n$ is shown in Fig.\,\ref{fig_phase_noise}\,(c), where the expected linear scaling of the timing jitter with $n$ is established by the linear fit of the data\,[see red dotted line in  Fig.\,\ref{fig_phase_noise}\,(c)]\,\cite{chatterjee2025comparative,razumov2023subspace}. 



\section{Linear Spectroscopy}
\label{sec:linear_spectroscopy}
As an example, we performed a proof-of-concept linear spectroscopy experiment with $ \Delta f = 2.5~\text{MHz}$. 
The experimental configuration follows the general architecture described in Subsection~\ref{sub_setup}. The output of the dual-comb light source is routed through a programmable waveshaper (Finisar Waveshaper 1000s) and down-converted through a fast PD (Thorlabs DETCFC08) and recorded using an oscilloscope. A Lorentzian spectral attenuation transfer function was programmed  into the waveshaper, with 5 dB maximum attenuation and 50 GHz full width at half maximum\,(FWHM), spanning five comb lines. Finally, the transfer function of the waveshaper was retrieved from the dual comb interferogram and compared with {\color{black} a reference measurement made with an optical spectrum analyzer\,(OSA), and the function programmed on the waveshaper.}
\begin{figure}[htb!]
    \centering
    \includegraphics[width=0.85\columnwidth]{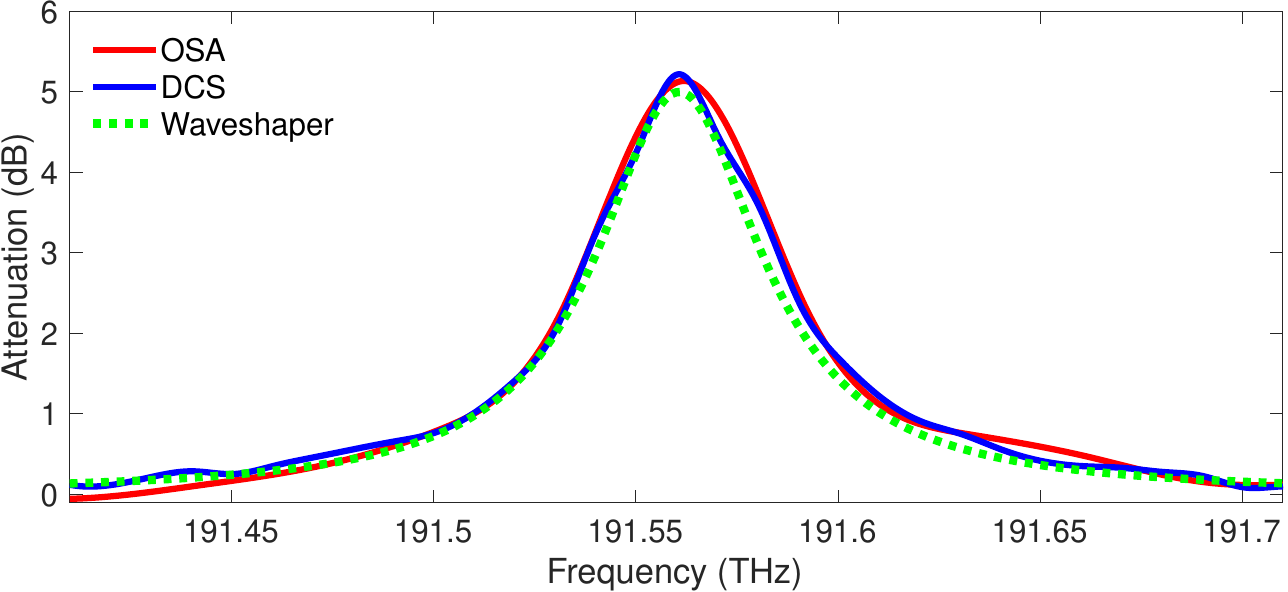}
    \caption{Linear spectroscopy validation of DCS light source using a programmable waveshaper. We plot measured waveshaper transfer function from DCS (solid blue line), measurement using an optical spectrum analyzer\,(OSA) (solid red line) and the programmed Lorentzian function in the waveshaper (dotted green line) with a 5 dB peak attenuation and 50 GHz FWHM.}
    \label{fig_linear_spec}
\end{figure}


The PSD of the reference interferogram {\color{black}(all frequencies transmitted through the waveshaper)} is subtracted from the measurement one {\color{black}(programmed waveshaper function on)}, and averaged over 20 recordings to retrieve the programmed transfer function. 
The obtained waveshaper transfer function is then 
compared with the {\color{black}OSA measurement and the} function programmed on the waveshaper, as shown in Fig.\,\ref{fig_linear_spec}. 
We see there is a good agreement between the {\color{black}OSA measurement,} programmed function and the DCS retrieved function of the waveshaper. 
This validates the potential of our system as an attractive platform for fast and broadband linear spectroscopy.

\section{Nonlinear Spectroscopy}\label{secNonLinSpec}
\begin{figure}[htb!]
\centering
\includegraphics[width=0.98\columnwidth]{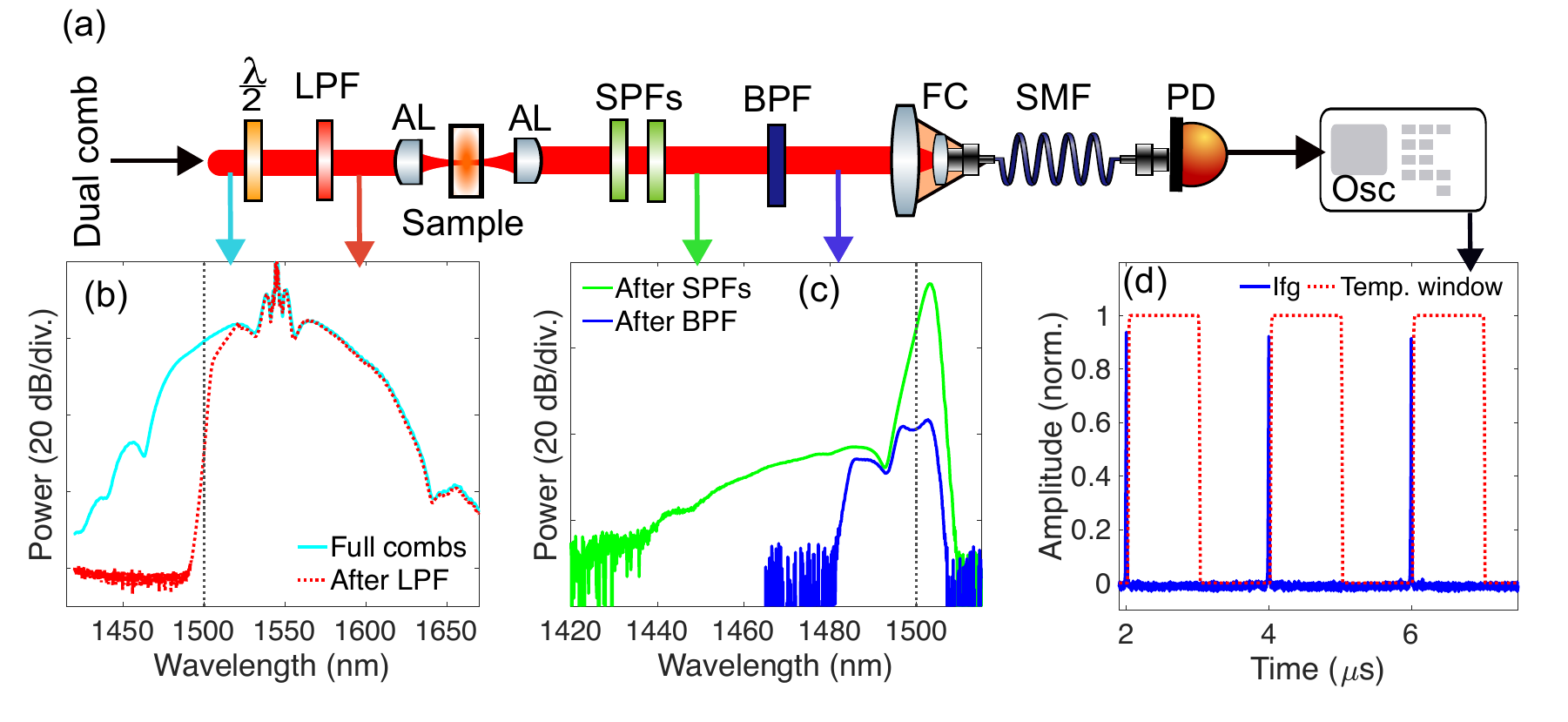}
\caption{(a) Schematic setup of dual comb CARS experiment. LPF : long wavelength pass filter, AL : aspheric lens, SPF : short wavelength pass filter, FC : fiber collimator, BPF : bandpass filter, SMF : single mode fiber, PD : photodetector, Osc : oscilloscope. (b,c) Optical spectra obtained at different positions of the setup from an optical spectrum analyzer. The cut on (off) wavelength of the filters, 1500 nm is indicated in grey dashed line for visual aid. Note that (b) and (c) have very different absolute optical powers. (d) Plot of acquired normalized time domain interferogram (blue) and a super-Gaussian filter (red) to eliminate the non-resonant background. Ifg : interferogram.}
\label{fig_carsbgosetup}
\end{figure}

Dual-comb CARS is an emerging nonlinear spectroscopy technique that aims to perform fast CARS measurements utilizing the speed-up attribute offered by DCS\,\cite{ideguchi2013coherent,mohler2017dual,kameyama2020dual,coluccelli2017fiber,lv2023ultrahigh,zhang2022delay,chen2017spectral,schweizer2026doppler}. In order to validate the utility of our dual-comb ultrashort light source, we demonstrated rapid retrieval of low-frequency Raman resonances ($<100$ cm$^{-1}$) in a {\color{black}1 mm thick}  Bi$_{12}$GeO$_{20}$ (BGO) crystal, at an interferogram acquisition rate $\Delta f=0.5$ MHz. {\color{black}Higher values of $\Delta f$ led to a degraded SNR of the detected CARS signal due to undersampling of the Raman oscillation.} The utilized setup is shown in Fig.\,\ref{fig_carsbgosetup}\,(a). 
The dual comb light emerging from the fiber beam combiner is launched into a long wavelength pass filter\,(LPF) with a cut on wavelength at 1500 nm. The LPF cuts out the part of the comb where we expect to see strong anti-Stokes Raman combs generated from the dual comb. The spectrum before and after the filter is recorded with an optical spectrum analyzer and shown in Fig.\,\ref{fig_carsbgosetup}\,(b). Then the beam is tightly focused (beam diameter $\sim$3 $\mu$m) onto the sample using C-coated aspheric lenses with 3 mm focal length. After passing through the sample, another aspheric lens is used to collimate back the light consisting of the launched pump and the generated four-wave mixing\,(FWM) combs consisting of the CARS light and the instantaneous non-resonant background. Then the light passes through a series of filters : two short wavelength pass filters\,(SPF) with cutoff wavelength 1550 nm and a bandpass filter\,(BPF) with 12 nm bandwidth centered at 1490 nm, to efficiently cut out the strong pump from reaching the PD\,\cite{mohler2017dual, carlson2020broadband}. The optical spectra recorded with an optical spectrum analyzer, after the SPFs and the BPF are shown in Fig.\,\ref{fig_carsbgosetup}\,(c). Note that in Fig.\,\ref{fig_carsbgosetup}\,(c), after the two SPFs, there is still some residual pump light forming the large peak around 1500 nm, with a much larger power than the broad FWM combs extending from 1430 nm to 1490 nm. Thus the BPF plays a crucial role in isolating the FWM generated light from the residual pump. Finally about 10 $\mu$W optical power of anti-Stokes light generated from the two combs was collected with a fiber-coupled collimator ($\sim$20\% coupling efficiency), and routed to an avalanche PD from Thorlabs with 400 MHz bandpass (PDB570C) to generate the interferogram signal. The interferogram was digitized with an oscilloscope at 1 GS/s sampling rate, and 500 traces were averaged. An example averaged interferogram trace is shown in blue solid line in Fig.\,\ref{fig_carsbgosetup}\,(d).

\begin{figure}[htb!]
\centering
\includegraphics[width=0.95\columnwidth]{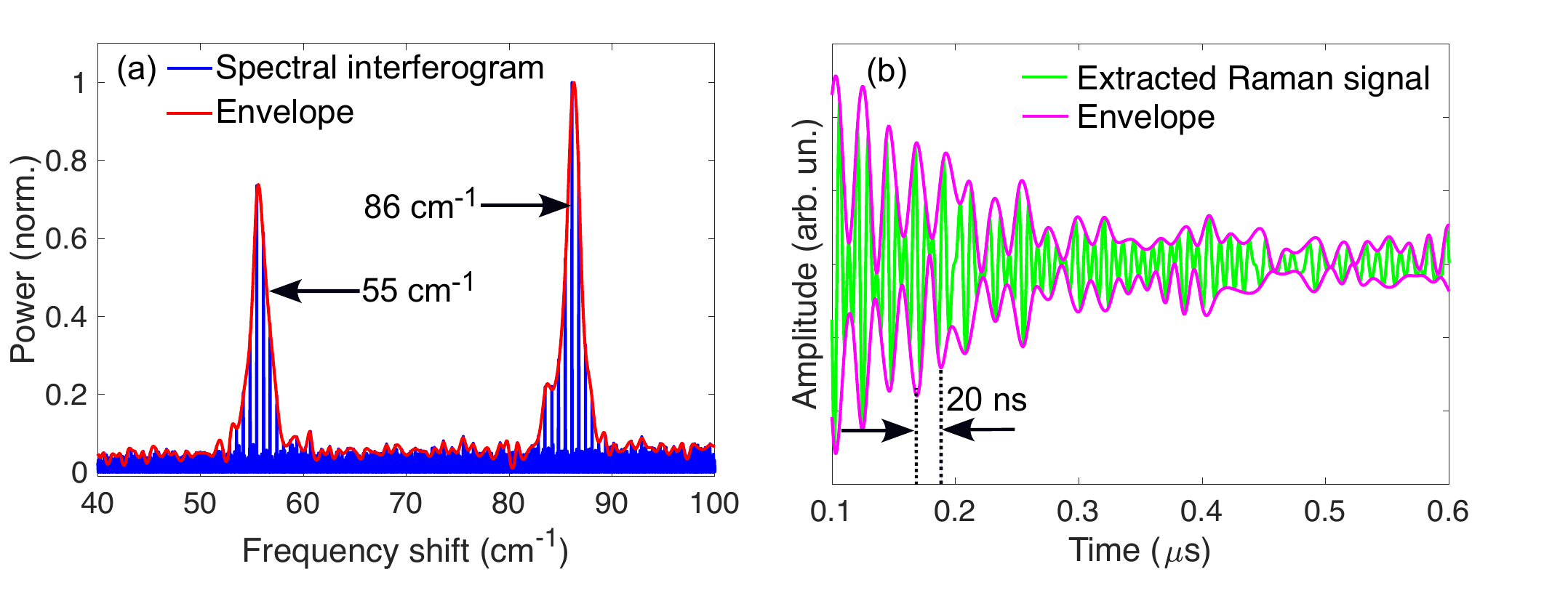}
\caption{(a) Frequency domain interferogram (blue) and its envelope (red) after temporal filtering and stretching the frequency axis by the magnification factor $m$. Two Raman modes at 55 cm$^{-1}$ and 86 cm$^{-1}$ are retrieved. (b) Time domain interferogram after temporal filtering. 20 ns time period beatnote is observed due to beating between the two Raman modes in the down-converted interferogram signal.}
\label{fig_carsresults}
\end{figure}

In order to remove the remaining non-resonant instantaneous background from the interferogram, in post-processing we employed a supergaussian temporal filter (order=80, duration=1\,$\mu$s) to isolate the part of the interferogram just after the centerburst\,\cite{junjuri2025review}. The temporal filter is shown in red dotted line in Fig.\,\ref{fig_carsbgosetup}\,(d). Then Fourier transforming the filtered time domain interferogram, and stretching the frequency axis by the magnification factor $m=20000$, sharp Raman modes of BGO near 55 cm$^{-1}$ and 86 cm$^{-1}$, were retrieved [see red envelope of down-converted comb lines in blue, in Fig.\,\ref{fig_carsresults}\,(a)], in agreement with the literature\,\cite{pylypets2025orbital,chen2011coherent}. An inverse Fourier transformed version of this signal (with the down-converted frequency axis) is shown in Fig.\,\ref{fig_carsresults}\,(b). In the time domain, the signal shows an exponential decay over about 100 ns, implying the Raman decoherence time of the strongest Raman mode to be $\frac{100\, \text{ns}}{m}=5$ ps. We also see a beatnote in the interferogram signal with a period of 20 ns, which arises due to the beating of the two Raman modes, that are separated by approximately 50 MHz in the down-converted frequency spectrum. 

In this experiment, the high interferogram acquisition rate of 0.5 MHz, and 500 averages make an effective refresh rate of our CARS spectroscopy to be 1 kHz. This illustrates the potential of EOM-based dual comb CARS systems for fast nonlinear spectroscopic imaging\,\cite{liu2026physics}.

\section{Conclusion \& Perspectives}\label{secCon}
To conclude, we demonstrated an all-PM fiber EOM-based broadband dual-comb light source delivering ultrashort (60 fs) pulses at 10 GHz repetition rate, {\color{black}where the combs were mixed  through an all-fiber beam combiner simplifying optical alignment. We highlighted its potential for fast linear and nonlinear spectroscopy through proof-of-concept experiments, making it attractive for advanced imaging applications requiring fast update rates\,\cite{chatterjee2025real,ideguchi2013coherent}. We also demonstrated that the use of an IQ modulator enables GHz-level tuning of the relative carrier-frequency shift between the two combs, allowing the RF comb to be optimally positioned within the Nyquist window. This, in turn, maximizes the accessible repetition-rate difference and minimizes the interferogram acquisition time while avoiding aliasing. Combining a 10-THz optical bandwidth with a repetition-rate difference as large as 2.5 MHz, our system achieves a figure of merit of 25 GHz$^2$, while performing linear spectroscopy, which to the best of our knowledge, is the highest reported for an alias-free EOM-based dual-comb system.} 
For nonlinear spectroscopy, low frequency ($<100$ cm$^{-1}$) Raman active modes of BGO were detected with a 0.5 MHz interferogram acquisition rate. {\color{black}{\color{black}Note that the agility of EOM comb source could further facilitate the} enhancement of the acquisition speed through pulse time-programming techniques\,\cite{lv2023ultrahigh, kameyama2020dual,zhang2022delay,wu2020repetition}.} These results open new perspectives in the development of nonlinear spectroscopic imaging modalities, especially relevant for tracking low-frequency Raman modes in semiconductor materials\,\cite{carlson2020broadband}, 2D materials\,\cite{zhao2013interlayer} and biological samples like viruses\,\cite{zhang2025nanoscopic}. 
{\color{black}Our system's all-PM fiber architecture, together with the fiber-based combiner that substantially simplifies optical alignment, paves the way towards a compact, alignment-free dual-comb CARS system for real-time imaging of a wide variety of samples.
}

\section{Acknowledgments}
Discussions with Francesco Tani and Siddharth Sivankutty are greatly appreciated. We thank Hicham El Hamzaoui for Raman characterization of the BGO sample. DC and AM acknowledge funding from the European Union, Marie Skłodowska-Curie Actions (MSCA) program through project FRESCOS. AM and HR acknowledge Agence Nationale de la Recherche (Programme Investissements d’Avenir) funding through project FARCO and COMBY. SB acknowledges the Hauts-de-France region for STARs funding via the DOMMiNo project. The authors acknowledge the support of the CDP C2EMPI, as well as the French State under the France-2030 programme, the University of Lille, the Initiative of Excellence of the University of Lille, the European Metropolis of Lille for their funding and support of the R-CDP-24-004-C2EMPI project. 

\section{Disclosures}
The authors declare no conflict of interest.

\section{Data Availability}
Data underlying the results presented in this paper are not publicly available at this time but may be obtained from the corresponding author upon reasonable request.

\bibliography{sample}

@article{rieker2014frequency,
  title={Frequency-comb-based remote sensing of greenhouse gases over kilometer air paths},
  author={Rieker, Gregory B and Giorgetta, Fabrizio R and Swann, William C and Kofler, Jon and Zolot, Alex M and Sinclair, Laura C and Baumann, Esther and Cromer, Christopher and Petron, Gabrielle and Sweeney, Colm and others},
  journal={Optica},
  volume={1},
  number={5},
  pages={290--298},
  year={2014},
  publisher={Optical Society of America}
}

@article{zeng2026integrated,
  title={Integrated Electro-Optic Frequency Combs: Physical Mechanisms, Device Architectures, Material Platforms and System Applications},
  author={Zeng, Hanqing and Hu, Qingyuan and Zhang, Yuebin and Liu, Xin and Zhuang, Yongyong and Wang, Zhihong and Wei, Xiaoyong and Xu, Zhuo},
  journal={Nanomaterials},
  volume={16},
  number={9},
  pages={559},
  year={2026},
  publisher={MDPI}
}

@article{wright2021two,
  title={Two-photon dual-comb LiDAR},
  author={Wright, Hollie and Sun, Jinghua and McKendrick, David and Weston, Nick and Reid, Derryck T},
  journal={Optics Express},
  volume={29},
  number={23},
  pages={37037--37047},
  year={2021},
  publisher={Optical Society of America}
}

@article{razumov2026impact,
  title={Impact of nonlinear spectral broadening on the phase noise properties of electro-optic frequency combs},
  author={Razumov, Aleksandr and Cai, Yijia and Stylios, Dimosthenis and Riebesehl, Jasper and Sillekens, Eric and Sohanpal, Ronit and Da Ros, Francesco and Liu, Zhixin and Zibar, Darko},
  journal={Optics Express},
  volume={34},
  number={4},
  pages={6229--6245},
  year={2026},
  publisher={Optica Publishing Group}
}

@article{chatterjee2025comparative,
  title={Comparative noise analysis of nonlinearly broadened dual frequency combs in different fiber propagation schemes},
  author={Chatterjee, Debanuj and Parriaux, Alexandre and Boivinet, Simon and Bouwmans, G{\'e}raud and Labat, Damien and Cassez, Andy and Mussot, Arnaud},
  journal={Optics Express},
  volume={33},
  number={21},
  pages={45014--45023},
  year={2025},
  publisher={Optica Publishing Group}
}

@article{chang2024dispersive,
  title={Dispersive Fourier transform based dual-comb ranging},
  author={Chang, Bing and Tan, Teng and Du, Junting and He, Xinyue and Liang, Yupei and Liu, Zihan and Wang, Chun and Xia, Handing and Wu, Zhaohui and Wang, Jindong and others},
  journal={Nature Communications},
  volume={15},
  number={1},
  pages={4990},
  year={2024},
  publisher={Nature Publishing Group UK London}
}

@article{nurnberg2021dual,
  title={Dual-comb ranging with frequency combs from single cavity free-running laser oscillators},
  author={N{\"u}rnberg, Jacob and Willenberg, Benjamin and Phillips, Christopher R and Keller, Ursula},
  journal={Optics Express},
  volume={29},
  number={16},
  pages={24910--24918},
  year={2021},
  publisher={Optical Society of America}
}

@article{zhu2015mid,
  title={Mid-infrared dual frequency comb spectroscopy based on fiber lasers for the detection of methane in ambient air},
  author={Zhu, F and Bicer, A and Askar, R and Bounds, J and Kolomenskii, AA and Kelessides, V and Amani, M and Schuessler, HA},
  journal={Laser Physics Letters},
  volume={12},
  number={9},
  pages={095701},
  year={2015},
  publisher={IOP Publishing}
}

@article{ideguchi2017dual,
  title={Dual-comb spectroscopy},
  author={Ideguchi, Takuro},
  journal={Optics and Photonics News},
  volume={28},
  number={1},
  pages={32--39},
  year={2017},
  publisher={OSA}
}

@article{chen2011coherent,
  title={Coherent optical phonon generation in Bi3Ge4O12},
  author={Chen, Z and Gao, Y and Minch, BC and DeCamp, MF},
  journal={Journal of Physics: Condensed Matter},
  volume={23},
  number={38},
  pages={385402},
  year={2011},
  publisher={IOP Publishing}
}

@article{pylypets2025orbital,
  title={Orbital angular momentum impact on light scattering by phonons tested experimentally},
  author={Pylypets, A and Borodavka, F and Rafalovskyi, I and Gregora, I and Buixaderas, E and Bohacek, P and Hlinka, J},
  journal={APL Photonics},
  volume={10},
  number={4},
  year={2025},
  publisher={AIP Publishing}
}

@article{chen2017spectral,
  title={Spectral focusing dual-comb coherent anti-Stokes Raman spectroscopic imaging},
  author={Chen, Kun and Wu, Tao and Chen, Tao and Wei, Haoyun and Yang, Honglei and Zhou, Tian and Li, Yan},
  journal={Optics Letters},
  volume={42},
  number={18},
  pages={3634--3637},
  year={2017},
  publisher={Optical Society of America}
}

@article{zhang2022delay,
  title={Delay-spectral focusing dual-comb coherent Raman spectroscopy for rapid detection in the high-wavenumber region},
  author={Zhang, Yujia and Lu, Minjian and Wu, Tao and Chen, Kun and Feng, Yongxiang and Wang, Wenhui and Li, Yan and Wei, Haoyun},
  journal={ACS Photonics},
  volume={9},
  number={4},
  pages={1385--1394},
  year={2022},
  publisher={ACS Publications}
}

@article{zhang2025nanoscopic,
  title={Nanoscopic acoustic vibrational dynamics of a single virus captured by ultrafast spectroscopy},
  author={Zhang, Yaqing and Wu, Rihan and Shahjahan, Md and Yang, Canchai and Pyeon, Dohun and Harel, Elad},
  journal={Proceedings of the National Academy of Sciences},
  volume={122},
  number={4},
  pages={e2420428122},
  year={2025},
  publisher={National Academy of Sciences}
}

@article{coluccelli2017fiber,
  title={Fiber-format dual-comb coherent Raman spectrometer},
  author={Coluccelli, Nicola and Howle, Christopher R and McEwan, Kenneth and Wang, Yuchen and Fernandez, Toney Teddy and Gambetta, Alessio and Laporta, Paolo and Galzerano, Gianluca},
  journal={Optics Letters},
  volume={42},
  number={22},
  pages={4683--4686},
  year={2017},
  publisher={Optical Society of America}
}

@article{lv2023ultrahigh,
  title={Ultrahigh-speed coherent anti-stokes Raman spectroscopy with a hybrid dual-comb source},
  author={Lv, Tianjian and Han, Bing and Yan, Ming and Wen, Zhaoyang and Huang, Kun and Yang, Kangwen and Zeng, Heping},
  journal={ACS Photonics},
  volume={10},
  number={8},
  pages={2964--2971},
  year={2023},
  publisher={ACS Publications}
}

@article{long2014multiheterodyne,
  title={Multiheterodyne spectroscopy with optical frequency combs generated from a continuous-wave laser},
  author={Long, David A and Fleisher, Adam J and Douglass, Kevin O and Maxwell, Stephen E and Bielska, K and Hodges, Joseph T and Plusquellic, David F},
  journal={Optics letters},
  volume={39},
  number={9},
  pages={2688--2690},
  year={2014},
  publisher={Optical Society of America}
}

@article{coburn2018regional,
  title={Regional trace-gas source attribution using a field-deployed dual frequency comb spectrometer},
  author={Coburn, Sean and Alden, Caroline B and Wright, Robert and Cossel, Kevin and Baumann, Esther and Truong, Gar-Wing and Giorgetta, Fabrizio and Sweeney, Colm and Newbury, Nathan R and Prasad, Kuldeep and others},
  journal={Optica},
  volume={5},
  number={4},
  pages={320--327},
  year={2018},
  publisher={Optical Society of America}
}

@inproceedings{liu2026physics,
  title={Physics and Applications of Dual-Comb Coherent Anti-Stokes Raman Spectroscopy for Biomedical Imaging},
  author={Liu, Bin and Wang, Jian and Xiuli, Luo and Xingcheng, Han and Hao, Gu},
  booktitle={Photonics},
  volume={13},
  number={2},
  pages={173},
  year={2026},
  organization={MDPI AG}
}

@article{boivinet2025polarization,
  title={Polarization maintaining multi-segment dispersion-tailored fibers for frequency comb generation with 27 fs pulses at 10 GHz repetition rate},
  author={Boivinet, S and Chatterjee, D and Houard, A and Kudlinski, A and Sivankutty, S and Conforti, M and Tani, Francesco and Mussot, A},
  journal={APL Photonics},
  volume={10},
  number={12},
  year={2025},
  publisher={AIP Publishing}
}

@article{merlin1997generating,
  title={Generating coherent THz phonons with light pulses},
  author={Merlin, R},
  journal={Solid state communications},
  volume={102},
  number={2-3},
  pages={207--220},
  year={1997},
  publisher={Elsevier}
}

@article{ideguchi2013coherent,
  title={Coherent Raman spectro-imaging with laser frequency combs},
  author={Ideguchi, Takuro and Holzner, Simon and Bernhardt, Birgitta and Guelachvili, Guy and Picqu{\'e}, Nathalie and H{\"a}nsch, Theodor W},
  journal={Nature},
  volume={502},
  number={7471},
  pages={355--358},
  year={2013},
  publisher={Nature Publishing Group UK London}
}

@article{mohler2017dual,
  title={Dual-comb coherent Raman spectroscopy with lasers of 1-GHz pulse repetition frequency},
  author={Mohler, Kathrin J and Bohn, Bernhard J and Yan, Ming and M{\'e}len, Gw{\'e}na{\"e}lle and H{\"a}nsch, Theodor W and Picqu{\'e}, Nathalie},
  journal={Optics letters},
  volume={42},
  number={2},
  pages={318--321},
  year={2017},
  publisher={Optical Society of America}
}

@article{chatterjee2025real,
  title={Real-time electro-optic dual comb detection of ultrasound waves},
  author={Chatterjee, Debanuj and Etien, Louis and Boivinet, Simon and Rigneault, Herv{\'e} and Chaigne, Thomas and Mussot, Arnaud},
  journal={Journal of Physics B: Atomic, Molecular and Optical Physics},
  volume={58},
  number={15},
  pages={153501},
  year={2025},
  publisher={IOP Publishing}
}

@article{okubo2015ultra,
  title={Ultra-broadband dual-comb spectroscopy across 1.0--1.9 $\mu$m},
  author={Okubo, Sho and Iwakuni, Kana and Inaba, Hajime and Hosaka, Kazumoto and Onae, Atsushi and Sasada, Hiroyuki and Hong, Feng-Lei},
  journal={Applied Physics Express},
  volume={8},
  number={8},
  pages={082402},
  year={2015},
  publisher={The Japan Society of Applied Physics}
}

@article{ishizawa2013phase,
  title={Phase-noise characteristics of a 25-GHz-spaced optical frequency comb based on a phase-and intensity-modulated laser},
  author={Ishizawa, Atsushi and Nishikawa, Tadashi and Mizutori, Akira and Takara, Hidehiko and Takada, Atsushi and Sogawa, Tetsuomi and Koga, Masafumi},
  journal={Optics express},
  volume={21},
  number={24},
  pages={29186--29194},
  year={2013},
  publisher={Optical Society of America}
}

@article{razumov2023subspace,
  title={Subspace tracking for phase noise source separation in frequency combs},
  author={Razumov, Aleksandr and Heeb{\o}ll, Holger R and Dummont, Mario and Terra, Osama and Dong, Bozhang and Riebesehl, Jasper and Varming, Poul and Pedersen, Jens E and Ros, Francesco Da and Bowers, John E and others},
  journal={Optics Express},
  volume={31},
  number={21},
  pages={34325--34347},
  year={2023},
  publisher={Optica Publishing Group}
}

@article{duran2015ultrafast,
  title={Ultrafast electrooptic dual-comb interferometry},
  author={Dur{\'a}n, Vicente and Tainta, Santiago and Torres-Company, Victor},
  journal={Optics Express},
  volume={23},
  number={23},
  pages={30557--30569},
  year={2015},
  publisher={Optical Society of America}
}

@article{newbury_sensitivity_2010,
	title = {Sensitivity of coherent dual-comb spectroscopy},
	volume = {18},
	copyright = {© 2010 OSA},
	issn = {1094-4087},
	url = {https://opg.optica.org/oe/abstract.cfm?uri=oe-18-8-7929},
	doi = {10.1364/OE.18.007929},
	number = {8},
	urldate = {2022-12-22},
	journal = {Optics Express},
	author = {Newbury, Nathan R. and Coddington, Ian and Swann, William},
	month = apr,
	year = {2010},
	pages = {7929--7945},
}

@article{ycas2018high,
  title={High-coherence mid-infrared dual-comb spectroscopy spanning 2.6 to 5.2 $\mu$m},
  author={Ycas, Gabriel and Giorgetta, Fabrizio R and Baumann, Esther and Coddington, Ian and Herman, Daniel and Diddams, Scott A and Newbury, Nathan R},
  journal={Nature Photonics},
  volume={12},
  number={4},
  pages={202--208},
  year={2018},
  publisher={Nature Publishing Group UK London}
}

@article{yan2017mid,
  title={Mid-infrared dual-comb spectroscopy with electro-optic modulators},
  author={Yan, Ming and Luo, Pei-Ling and Iwakuni, Kana and Millot, Guy and H{\"a}nsch, Theodor W and Picqu{\'e}, Nathalie},
  journal={Light: Science \& Applications},
  volume={6},
  number={10},
  pages={e17076--e17076},
  year={2017},
  publisher={Nature Publishing Group}
}

@article{long_nanosecond_2024,
	title = {Nanosecond time-resolved dual-comb absorption spectroscopy},
	volume = {18},
	copyright = {2023 This is a U.S. Government work and not under copyright protection in the US; foreign copyright protection may apply},
	issn = {1749-4893},
	url = {https://www.nature.com/articles/s41566-023-01316-8},
	doi = {10.1038/s41566-023-01316-8},
	language = {en},
	number = {2},
	urldate = {2024-06-27},
	journal = {Nature Photonics},
	author = {Long, David A. and Cich, Matthew J. and Mathurin, Carl and Heiniger, Adam T. and Mathews, Garrett C. and Frymire, Augustine and Rieker, Gregory B.},
	month = feb,
	year = {2024},
	note = {Publisher: Nature Publishing Group},
	pages = {127--131},
}

@article{hoghooghi_broadband_2022,
	title = {Broadband 1-{GHz} mid-infrared frequency comb},
	volume = {11},
	copyright = {2022 The Author(s)},
	issn = {2047-7538},
	url = {https://www.nature.com/articles/s41377-022-00947-w},
	doi = {10.1038/s41377-022-00947-w},
	language = {en},
	number = {1},
	urldate = {2024-06-27},
	journal = {Light: Science \& Applications},
	author = {Hoghooghi, Nazanin and Xing, Sida and Chang, Peter and Lesko, Daniel and Lind, Alexander and Rieker, Greg and Diddams, Scott},
	month = sep,
	year = {2022},
	note = {Publisher: Nature Publishing Group},
	pages = {264},
}

@article{coddington_dual-comb_2016,
	title = {Dual-comb spectroscopy},
	volume = {3},
	issn = {2334-2536},
	url = {https://opg.optica.org/optica/abstract.cfm?uri=optica-3-4-414},
	doi = {10.1364/OPTICA.3.000414},
	language = {EN},
	number = {4},
	urldate = {2024-06-27},
	journal = {Optica},
	author = {Coddington, Ian and Newbury, Nathan and Swann, William},
	month = apr,
	year = {2016},
	note = {Publisher: Optica Publishing Group},
	pages = {414--426},
}

@article{schliesser2005frequency,
  title={Frequency-comb infrared spectrometer for rapid, remote chemical sensing},
  author={Schliesser, Albert and Brehm, Markus and Keilmann, Fritz and van der Weide, Daniel W},
  journal={Optics express},
  volume={13},
  number={22},
  pages={9029--9038},
  year={2005},
  publisher={Optica Publishing Group}
}

@article{millot2016frequency,
  title={Frequency-agile dual-comb spectroscopy},
  author={Millot, Guy and Pitois, St{\'e}phane and Yan, Ming and Hovhannisyan, Tatevik and Bendahmane, Abdelkrim and H{\"a}nsch, Theodor W and Picqu{\'e}, Nathalie},
  journal={Nature Photonics},
  volume={10},
  number={1},
  pages={27--30},
  year={2016},
  publisher={Nature Publishing Group UK London}
}

@article{wu2020repetition,
  title={Repetition frequency modulated fiber laser for coherent anti-Stokes Raman scattering},
  author={Wu, Tao and Chen, Kun and Wei, Haoyun and Li, Yan},
  journal={Optics Letters},
  volume={45},
  number={2},
  pages={407--410},
  year={2020},
  publisher={Optical Society of America}
}

@article{coddington2008coherent,
  title={Coherent multiheterodyne spectroscopy using stabilized optical frequency combs},
  author={Coddington, Ian and Swann, William C and Newbury, Nathan R},
  journal={Physical Review Letters},
  volume={100},
  number={1},
  pages={013902},
  year={2008},
  publisher={APS}
}

@article{schweizer2026doppler,
  title={Doppler dual-comb coherent Raman spectromicroscopy},
  author={Schweizer, Florian M and Terrasa, Hannah and Garg, Manish},
  journal={arXiv preprint arXiv:2603.23094},
  year={2026}
}

@article{junjuri2025review,
  title={Review of Coherent Anti-Stokes Raman Scattering Nonresonant Background Removal and Phase Retrieval Approaches: From Experimental Methods to Deep Learning Algorithms},
  author={Junjuri, Rajendhar and Bocklitz, Thomas},
  journal={Advanced Photonics Research},
  volume={6},
  number={9},
  pages={2500035},
  year={2025},
  publisher={Wiley Online Library}
}

@article{zhao2013interlayer,
  title={Interlayer breathing and shear modes in few-trilayer MoS2 and WSe2},
  author={Zhao, Yanyuan and Luo, Xin and Li, Hai and Zhang, Jun and Araujo, Paulo T and Gan, Chee Kwan and Wu, Jumiati and Zhang, Hua and Quek, Su Ying and Dresselhaus, Mildred S and others},
  journal={Nano letters},
  volume={13},
  number={3},
  pages={1007--1015},
  year={2013},
  publisher={ACS Publications}
}

@article{nishiyama2017sensitivity,
  title={Sensitivity improvement of dual-comb spectroscopy using mode-filtering technique},
  author={Nishiyama, Akiko and Yoshida, Satoru and Hariki, Takuya and Nakajima, Yoshiaki and Minoshima, Kaoru},
  journal={Optics express},
  volume={25},
  number={25},
  pages={31730--31738},
  year={2017},
  publisher={Optica Publishing Group}
}

@article{carlson2020broadband,
  title={Broadband, electro-optic, dual-comb spectrometer for linear and nonlinear measurements},
  author={Carlson, David R and Hickstein, Daniel D and Papp, Scott B},
  journal={Optics Express},
  volume={28},
  number={20},
  pages={29148--29154},
  year={2020},
  publisher={Optica Publishing Group}
}

@article{parriaux2020electro,
  title={Electro-optic frequency combs},
  author={Parriaux, Alexandre and Hammani, Kamal and Millot, Guy},
  journal={Advances in Optics and Photonics},
  volume={12},
  number={1},
  pages={223--287},
  year={2020},
  publisher={Optica Publishing Group}
}

@article{kameyama2020dual,
  title={Dual-comb coherent Raman spectroscopy with near 100\% duty cycle},
  author={Kameyama, Risako and Takizawa, Shigekazu and Hiramatsu, Kotaro and Goda, Keisuke},
  journal={ACS Photonics},
  volume={8},
  number={4},
  pages={975--981},
  year={2020},
  publisher={ACS Publications}
}

@article{yu2021phase,
  title={Phase-stable repetition rate multiplication of dual-comb spectroscopy based on a cascaded Mach--Zehnder interferometer},
  author={Yu, Haoyang and Qian, Zhou and Xinghui, Li and Wang, Xiaohao and Ni, Kai},
  journal={Optics Letters},
  volume={46},
  number={13},
  pages={3243--3246},
  year={2021},
  publisher={Optica Publishing Group}
}

@article{caldwell2022time,
  title={The time-programmable frequency comb and its use in quantum-limited ranging},
  author={Caldwell, Emily D and Sinclair, Laura C and Newbury, Nathan R and Deschenes, Jean-Daniel},
  journal={Nature},
  volume={610},
  number={7933},
  pages={667--673},
  year={2022},
  publisher={Nature Publishing Group UK London}
}

@article{voumard20221,
  title={1-GHz dual-comb spectrometer with high mutual coherence for fast and broadband measurements},
  author={Voumard, Thibault and Darvill, John and Wildi, Thibault and Ludwig, Markus and Mohr, Christian and Hartl, Ingmar and Herr, Tobias},
  journal={Optics letters},
  volume={47},
  number={6},
  pages={1379--1382},
  year={2022},
  publisher={Optica Publishing Group}
}

@article{hoghooghi202111,
  title={11-$\mu$s time-resolved, continuous dual-comb spectroscopy with spectrally filtered mode-locked frequency combs},
  author={Hoghooghi, Nazanin and Cole, Ryan K and Rieker, Gregory B},
  journal={Applied Physics B},
  volume={127},
  number={2},
  pages={17},
  year={2021},
  publisher={Springer}
}

@article{giorgetta2024broadband,
  title={Broadband dual-comb hyperspectral imaging and adaptable spectroscopy with programmable frequency combs},
  author={Giorgetta, Fabrizio R and Desch{\^e}nes, Jean-Daniel and Lieber, Richard L and Coddington, Ian and Newbury, Nathan R and Baumann, Esther},
  journal={APL Photonics},
  volume={9},
  number={1},
  year={2024},
  publisher={AIP Publishing}
}

@article{chatterjee2025sensitivity,
  title={Sensitivity Improvement in Dual Comb Spectroscopy With Time-Programmed Electro-Optic Frequency Combs},
  author={Chatterjee, Debanuj and Bancel, Eve-Line and Conforti, Matteo and Sivankutty, Siddharth and Rigneault, Herv{\'e} and Szriftgiser, Pascal and Cundiff, Steven and Mussot, Arnaud},
  journal={Journal of Lightwave Technology},
  volume={43},
  number={10},
pages     = {4648 -- 4658},
  year={2025},
  publisher={IEEE}
}



\end{document}